\documentstyle[epsfig]{aipproc}
\def\bea{\begin{eqnarray}}
\def\eea{\end{eqnarray}}
\def\be{\begin{equation}}
\def\ee{\end{equation}}
\newcommand{\nn}{\nonumber}
\begin{document}
\title{The impact of atomic precision measurements in high energy physics
\footnote{
Talk given at the XVII International Conference on Atomic
Physics, ICAP 2000, Florence, June 4-9, 2000}}

\author{Roberto Casalbuoni}
\address{Dept. of Physics, University of Florence and I.N.F.N., Florence\\
e-mail: casalbuoni@fi.infn.it}

\maketitle

\begin{abstract}
In this talk I discuss the relevance of atomic physics in
understanding some important questions about elementary particle
physics. A particular attention is devoted to atomic parity
violation measurements which seem to suggest new physics beyond
the Standard Model. Atomic physics might also be relevant in
discovering possible violations of the CPT symmetry.
\end{abstract}

\section*{Introduction}

The aim of this talk is to review some of the atomic precision
measurements in atomic physics leading to precious informations
in the realm of high-energy physics. The idea of atomic physics
bringing light on the high-energy physics world requires some
qualification due to the very different scales of energy involved
in the two cases. In fact, typically one has a separation of
about six or seven order of magnitude between the two scales and
one expects the two physics being almost decoupled. In fact, if we
look at some observable, $A$, at a scale $\Lambda_1\ll\Lambda_2$,
we expect that the observable can be represented in the form
\begin{equation}
A(\Lambda_1,\Lambda_2)=A(\Lambda_1)+{\cal
O}\left(\left(\frac{\Lambda_1}{\Lambda_2}\right)^n\right)
\end{equation}
In order to be able to derive informations about the physics at
the scale $\Lambda_2$, being at the scale $\Lambda_1$, one starts
considering a combination of observables corresponding to the
corrections coming from the higher scale
\begin{equation}
B=c\left(\frac{\Lambda_1}{\Lambda_2}\right)^n
\end{equation}
In order to measure $B$ one needs either  the coefficient $c$
being very large in such a way to partially compensate the scale
factor, or having an extremely good experimental sensitivity. In
this talk I will consider two particular examples of situations
where atomic physics can be relevant to high-energy physics,
namely Atomic Violation of Parity (APV) and possible violations
of the discrete symmetry CPT, that is the product of
charge-conjugation, parity and time-reversal. In fact, using
heavy atoms like cesium in APV measurements one can get a good
enhancement factor. On the other hand, CPT symmetry can be tested
by using the extraordinary opportunities offered by the atomic
traps in order to obtain a very accurate determination   of
frequencies.

\section*{Atomic Parity Violation in Atoms}

In this Section I will discuss mainly the latest determination of
the weak charge in atomic cesium and some of its implications in
models of physics beyond the Standard Model (SM).
The SM has been tested very precisely at machines such as LEP and
SLC, where, working at an energy around the $Z$ mass, one is
mainly testing the property of the $Z$ itself. Therefore, the
physics beyond the SM that can be looked for at these machines is
the one giving corrections to the $Z$-propagator and/or to the
couplings of the $Z$ with  fermion-antifermion pairs. Namely, new
massive vector bosons, $Z'$, which mix to the $Z$, or new
particles running in loops and  contributing to the $Z$
self-energy or to vertex corrections. But consider, for instance,
the case of a massive vector boson which does not mix to the $Z$,
and therefore invisible at LEP (except for tiny radiative
corrections). If the $Z'$ is coupled to fermions, in the
low-energy limit  it gives rise to an effective four-fermi
interaction.
Therefore, low-energy experiments are complementary to the
high-energy ones, and furthermore they are able to measure
directly the couplings of the $Z$ to light quarks; something that
at LEP and SLC can be done only in an indirect way. Among the
low-energy experiments a particular role is played by the APV
experiments, due to the precision almost at the level of the one
reached at LEP/SLC.

Let us now recall some feature of  APV in atoms. First of all,
within the SM the four-fermi parity violating hamiltonian density
for  nucleons is given by
\begin{equation}
{\cal H}^{PV}=\frac{G_F}{\sqrt{2}}\left[(\bar e\gamma_\mu\gamma_5
e)\sum_{N=p,n}c_{1N}\bar N\gamma^\mu N+(\bar e\gamma_\mu
e)\sum_{N=p,n}c_{2N}\bar N\gamma^\mu\gamma_5 N\right]
\end{equation}
where \be
c_{ip}=-2c_{iu}-c_{id},~~~c_{in}=-c_{iu}-2c_{id},~~~i=1,2\ee and
\be c_{1q}=-8a_ev_q=-(T_3^q-2s_\theta^2Q^q),~~~c_{2q}=-8v_ea_q=
-T_3^q(1-4s_\theta^2),~~~q=u,d\ee Here $v_e$, $v_q$, $a_e$ and
$a_q$ are the vector and vector-axial couplings of the $Z$ to the
electrons and quarks. For a point-like nucleus with $Z$ protons
and $N$ neutrons, the hamiltonian density, in the non-relativistic
limit, is given by \bea {\cal
H}_{PV}&=&\frac{G_F}{4\sqrt{2}m_{e}}\Big[\,Q_W(Z,N)
\vec\sigma_\ell\cdot[\vec p\,,\delta^3(\vec r)]_+ + 2(c_{2p}\vec
S_p+ c_{2n}\vec S_n)\cdot[\vec p\,,\delta^3(\vec r)]_+\\&-&
2i\vec\sigma_\ell\wedge(c_{2p}\vec S_p+c_{2n}\vec S_n)\cdot [\vec
p\,,\delta^3(\vec r)]_+\Big] \eea where $\vec p$ is the momentum
of the electron, $\vec S_{p(n)}$  the total spin of the protons
(neutrons) and $m_e$ the electron mass. I have also defined the
{\it weak charge} of the atom as\be
Q_W(Z,N)=2\left[c_{1p}Z+c_{1n}N\right]\ee Notice that for a heavy
atom (large values of $Z$) the matrix element of the first term
in ${\cal H}_{PV}$ is roughly proportional to $Z^3$, one factor
coming from $Q_W$, one from the momentum of the electron and the
third one from the wave function evaluated at the origin. This
coherence effect was noticed by Bouchiat and Bouchiat
\cite{bouchiat2} and it provides, in the case of cesium ($Z=55$)
an enhancement factor of about $10^5$, more or less what is
necessary in order  to compensate for the decoupling factor from
the scales mentioned in the Introduction.

In order to get a rough idea of the bounds on new physics that
can be obtained by a measurement of $Q_W$ with a given
sensitivity, we parametrize the new physics contribution to $Q_W$
by a four-fermi effective interaction \cite{musolf}\be {\cal
L}_{NP}^{PV}= \frac{g^2_{NP}}{\Lambda^2}\,\bar e
\gamma_\mu\gamma_5\, e\sum_{q=u,d}h_{1q\,}\bar q\, \gamma^\mu\
q\ee If we assume $h_{1q}\approx c_{1q}$, for a sensitivity
$\Delta Q_W/Q_W\approx 1\%$ one gets a bound \be\Lambda\approx
(5\, g_{NP})~TeV\ee If new physics is strongly interacting
($g_{NP}^2\approx 4\pi$), then $\Lambda\approx 17~TeV$, whereas
in the weakly interacting case ($g_{NP}^2\approx 4\pi\alpha$) we
get $\Lambda\approx 1.5~TeV$. In any case we see that at 1\%
level of  sensitivity, $Q_W$ is able to test new physics for
scales greater than 1 $TeV$.

In APV measurements  one looks at optical transitions between a
pair of states $|\psi_\pm\rangle$ mixed by ${\cal H}_{PV}$ and a
state $|\psi_0\rangle$ of the same nominal parity as
$|\psi_+\rangle$. The mixing of the two eigenstates of parity is
given by \be
\eta=\frac{\langle\psi_-|H_{PV}|\psi_+\rangle}{\Delta E}\ee where
$\Delta E$ is the splitting between the two levels. If I denote by
$M_1$ and $E_1^{PV}$ the amplitudes for the two unperturbed
transitions $|\psi_+\rangle\to|\psi_0\rangle$ and
$|\psi_-\rangle\to|\psi_0\rangle$,  the transition probability,
after the mixing, is given by \be W=M_1^2+|E_1^{PV}|^2\pm 2
\,Im\,(E_1^{PV})M_1\ee The choice of the sign depends on the
helicity of the photon which is emitted or absorbed in the
transition. In the actual experiment on cesium one measures the
{\it circular dichroism}, that is the asymmetry for the absorption
cross-section \be \delta=\frac{\sigma_+-\sigma_-}{\sigma_+
+\sigma_-}\approx 2\, \frac{Im\,(E_1^{PV})}{M_1}\ee Of course, the
PV amplitude $E_1^{PV}$ is proportional to the mixing parameter
$\eta$ and therefore measuring $\delta$ one can get the matrix
element of the PV hamiltonian. These ideas have been applied in
particular to the transition $6S\to 7S$ in atomic cesium
$^{133}_{55}Cs$ \cite{bouchiat,boulder1,boulder2}, but also to
other atoms as thallium \cite{thallium}. The typical value of
$\delta$ is $10^{-4}\div 10^{-5}$, but there is a strong
background which can be overcomed by letting the PV amplitude to
interfere with a large electro-induced (Stark) transition.
Eventually one extracts from the experiment the matrix element of
$H^{PV}$ which is proportional to $Q_W$ times an atomic form
factor $\kappa_{PV}$ which must be evaluated theoretically in
order to extract the value of the weak charge. Therefore the
measurement must be coupled with theoretical calculations of
similar accuracy in order to get a precise determination of
$Q_W$. In the case of atomic cesium the calculation of
$\kappa_{PV}$ was performed independently by two groups
\cite{dzuba,blundell}.  This calculation  is not an easy task, as
one has to use many-body perturbation theory coupled with
Hartree-Fock techniques. The theoretical errors are quite
difficult to  estimate. The authors of Refs.
\cite{dzuba,blundell} did their estimate by looking at the
differences between  the theoretical and the experimental values
of parity conserving quantities as dipole matrix elements and
hyperfine splittings for the $6S_{1/2}$, $7S_{1/2}$, $6P_{1/2}$
and $7P_{1/2}$ states. In this way the error
$\Delta\kappa_{PV}/\kappa_{PV}\approx 1\%$ was obtained. After
the new measurement of the weak charge of the cesium by the
Boulder group \cite{boulder2}, which improved the accuracy of the
previous experiment \cite{boulder1} by more than a factor five,
Bennett and Wieman \cite{bennett} re-examined the theoretical
errors on $\kappa_{PV}$. In fact, since the time of the previous
estimate there have been a number of new and more precise
measurements of the quantities of interest. The result is that
now the agreement is much better than before, and as a
consequence Bennett and Wieman got the estimate
$\Delta\kappa_{PV}/\kappa_{PV}\approx 0.4\%$.  It should be
noticed that there is a third element which contributes to the
extraction of $Q_W$ from the data. This is the Stark
mixing-induced electric dipole moment amplitude, $\beta$. The
experiments in Refs. \cite{bouchiat,boulder1} were using a
theoretical determination of $\beta$. In \cite{boulder2} the ratio
$M_{hf}/\beta$ has been measured. The off-diagonal magnetic
dipole moment induced by the hyperfine interaction is well known
empirically and it is possible to extract a precise value for
$\beta$. However, in a contribution to this Conference
\cite{dzuba2}, the matrix element $M_{hf}$ has been accurately
calculated with the result that the empirical formula for it
should be corrected by a factor of  0.24\% increasing the
discrepancy with the SM (see later). I would like also to comment
about some possible neglected contribution in the evaluation of
the atomic form factor. It has been pointed out in ref.
\cite{pollock} that there could be a contribution arising from
the difference of neutron and proton spatial distributions inside
the nucleus. This contribution turns out to be very difficult to
estimate, in fact it is quite model dependent. Most probably it
could introduce a further error on $Q_W(Cs)$ of about 0.3. This
would not change the conclusions in a very significant way.
Another point has been raised recently in ref. \cite{derevianko}.
This author argues that the contribution from the Breit
interaction (exchange of a transverse photon between two
electrons) could have been underestimated. The Breit interaction
contribution to the atomic form factor was estimated in
\cite{dzuba} and it was found to be very small. However in ref.
\cite{derevianko}  it is found that the total effect, taking into
account also second and third order contributions, is about twice
the first order effect. As a consequence, if "all" the higher
order contributions could be shown to be negligible, the
experimental measure would reconcile with the SM expectation for
$Q_W(Cs)$. However, see also ref. \cite{kozlov}.

To conclude this analysis I think that an evaluation of the atomic
form factor by taking into account the next order in the many-body
perturbative theory is highly desirable in order to settle the
question. In any case I find of some interest to
{\underbar{assume}} that the theoretical error is indeed at the
level of $0.4\%$ in order to see which are the possible
implications of the APV in high-energy physics.

I can start now to discuss the experimental results on $Q_W(Cs)$.
It is interesting to recall the value obtained in \cite{boulder1}
combined with the theoretical determination of $\kappa_{PV}$
\cite{dzuba,blundell} \be Q_W(Cs)=-71.04\pm(1.58)_{\rm exp}\pm
(0.88)_{\rm th}\ee The total error of these measurement on
$Q_W(Cs)$ is at  2.5\% level of accuracy  that, at that time, was
comparable with the sensitivity obtained at LEP1. In fact, this
determination of $Q_W(Cs)$ lead to the first indication that
technicolor models, in their most simple version obtained from
scaling of QCD, could not possibly fit the data. The new
experimental result on $Q_W(Cs)$ \cite{boulder2} combined with
the new determination of the theoretical error \cite{bennett}
gives \be Q_W(Cs)^{\rm exp} =-72.06\pm(0.28)_{\rm exp}\pm
(0.34)_{\rm th}\label{qwexp}\ee A result at 0.6\% level of
accuracy. On the theoretical side, $Q_W$ can be expressed as
\cite{marciano} \be Q_W(Cs)^{\rm th}=-72.72\pm
0.13-102\epsilon_3^{\rm rad}+\delta_NQ_W \label{qw}\ee including
hadronic-loop uncertainty. I use here the variables $\epsilon_i$
(i=1,2,3) of ref. \cite{altarelli}, which include the radiative
corrections, in place of the set of variables $S$, $T$ and $U$
originally introduced in ref. \cite{peskin}. In the above
definition of $Q_W^{\rm th}(Cs)$ I have explicitly included only
the Standard Model (SM) contribution to the radiative corrections.
New physics (that is physics beyond the SM) contributions are
represented by the term $\delta_N Q_W$. Also, I have neglected a
correction proportional to $\epsilon_1^{\rm rad}$. In fact, as
well known \cite{marciano}, due to the particular values of the
number of neutrons ($N=78$) and  protons ($Z=55$) in cesium, the
dependence on $\epsilon_1$ almost cancels out. For a top mass of
175 $GeV$ and $m_H=100(300)~GeV$ the value of $\epsilon_3^{\rm
rad}$ is given by \cite{altarelli2}\be \epsilon_3^{\rm
rad}=5.110(6.115)\times 10^{-3}\ee For $m_H=100~GeV$,
corresponding roughly to the lower experimental bound from direct
search at LEP2 \cite{higgs}, one gets \be Q_W(Cs)^{\rm
exp}-Q_W(Cs)^{SM}=1.18\pm 0.46 \label{bounds}\ee giving rise  to a
deviation of about $2.57~SD$. Furthermore, for increasing mass of
the Higgs the discrepancy increases. Therefore, if we assume as
being correct  the experimental result, the theoretical evaluation
of $\kappa_{PV}$ and the evaluation of the theoretical errors, we
are forced to conclude that the {\underbar {SM is disfavored at
99\% CL}}.

We can draw another conclusion, that is,  that in order to explain
the data on $Q_W(Cs)$ we need new physics not constrained by the
LEP and SLC data. In fact, as an example let me consider a type of
new physics visible at LEP as, for instance, contributing to the
self-energy of the $Z$, the so called oblique corrections. In such
a case one can write $\delta_NQ_W({\rm
oblique})=-102\epsilon_{3N}$, and in order to compensate for the
discrepancy on $Q_W(Cs)$ one  needs \be
\epsilon_{3N}=(-11.6\pm4.5)\times 10^{-3}\ee whereas from LEP and
SLC data one can determine the sum \be\epsilon_3^{\rm
exp}=\epsilon_3^{\rm rad}+\epsilon_{3N}=(4.19\pm 1)\times
10^{-3}\ee Therefore one gets $\epsilon_{3N}\approx 10^{-3}$, one
order of magnitude too small to explain the data on $Q_W(Cs)$.

I would like also recall the experimental result of APV on
Thallium \cite{thallium} \be Q_W(Tl)^{\rm exp}=-114.8\pm (1.2)
_{\rm exp}\pm (3.4)_{\rm th}\ee This result is not as precise as
the one on $Cs$, and in fact the total error is about 3\%. At this
level it is perfectly compatible with the SM prediction \be
Q_W(Tl)^{\rm SM}=-116.7\pm 0.1\ee

A new experiment on cesium is being planned in Paris but the
experimental sensitivity is going to be lower than the one
obtained in Boulder.

In Berkeley and Seattle there are plans for isotope ratio
measurements. In this case the dependence on the atomic form
factor would go away eliminating the theoretical error. However
these ratios depend on the variation of the neutron density along
the isotope chain. This would introduce errors at least twice as
big as the experimental ones \cite{chen}.

We are now in the position of discussing the implications of eq.
(\ref{bounds}) on new physics. Assuming that the contribution of
new physics, $\delta_NQ_W$, is such to reproduce the experimental
results, we can make use of eqs. (\ref{qwexp}) and (\ref{qw}) to
write \cite{casalbuoni}\be Q_W(Cs)^{\rm exp}-Q_W(Cs)^{\rm
th}(m_H)=0.66+102\epsilon_3^{\rm rad}(m_H)-\delta_NQ_W\pm0.46\ee
For $m_H=100~GeV$ at a 95\% CL we find \be 0.28\le\delta_NQ_W\le
2.08\ee Notice that the lower positive bound arises since the SM
(corresponding to $\delta_NQ_W=0$) does not fit the experimental
value of $Q_W(Cs)$ at this CL value. This is quite important
since it implies an {\underbar{upper bound}} on the scale of new
physics. For the same reason new physics with a contribution
$\delta_NQ_W<0$ is not allowed. Also notice that lower and upper
bounds both increase for increasing Higgs mass.
\medskip
\paragraph*{Contact interactions from compositness.}
A typical four-fermi operator in composite models contributing to
the PV lagrangian is \cite{langacker,casalbuoni} \be \pm
\frac{g^2}{\Lambda^2}\,\bar e\, \gamma_\mu\frac{1-\gamma_5}2\,
e\bar q\,\gamma^\mu\frac{1-\gamma_5}2\,q \label{contact}\ee The
effect of this interaction is to modify the coefficients
$c_{1u,1d}$ \be c_{1u,1d}\to c_{1u,1d}\mp
\frac{\sqrt{2}\pi}{G_F\Lambda^2}\ee  where, since composite
models correspond to strongly interacting new physics, we have
assumed $g^2=4\pi$. From \be Q_W=-2[(2Z+N)c_{1u}+(Z+2N)c_{1d}]\ee
we see that the negative sign for the operator (\ref{contact}) is
excluded. For the positive sign we get the bounds \be
12.1\le\Lambda(TeV)\le 32.9\ee The typical lower bound from high
energy physics is about 3.5 $TeV$ \cite{PDG}.
\medskip
\paragraph*{Extra-dimension models.}
In ref. \cite{pomarol} a minimal  extension  to higher dimensions
of the SM, with extra dimensions compactified, was considered. In
this model the fermions live in a 4-dimensional subspace, the
wall, whereas the gauge bosons live in the full D-dimensional
space, the bulk. In general, there might be two Higgs fields, one
living in the bulk, $\phi_1$, and the other living on the wall,
$\phi_2$. The propagation of the gauge fields in the bulk is
equivalent to the exchange of an infinite tower of Kaluza-Klein
(KK) excitations with increasing mass. For example, for $D=5$,
$M=n/R$, $n=1,\cdots,\infty$, with $R$ the compactification
radius. If  only the Higgs field $\phi_2$ is present, the
ordinary gauge bosons do not mix with the KK resonances and it is
easy to see that the contribution of these modes to $Q_W$ is
negative \cite{casalbuoni2}. Therefore the model does not fit the
data on $Q_W(Cs)$. For the more general case of both Higgs fields
present it has been shown \cite{casalbuoni2} that the LEP/SLC and
$Q_W(Cs)$ experimental data are not compatible among them at 95\%
CL.
\medskip
\paragraph*{Extra $Z'$ models.}

\begin{figure}[t!] 
\centerline{\epsfig{file=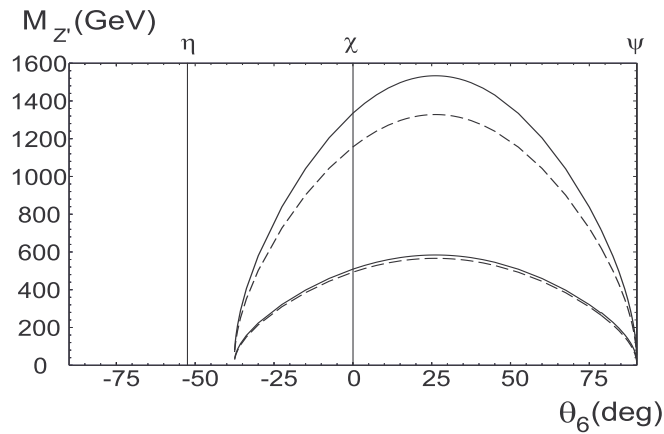,height=6cm,width=9cm}}
\vspace{10pt} \caption{The Figure shows the 95\% CL regions
allowed by $Q_W$ for the $Z'$ models. The solid contour
corresponds to $m_H=100~GeV$, and the dashed one to
$m_H=300~GeV$.} \label{zetap}
\end{figure}

The implications of models with an extra neutral vector boson
$Z'$ for APV have been considered in the literature for quite a
long time \cite{qwz,casalbuoni2,erler}. The $Z'$ has couplings
comparable to the ones of the $Z$ in the SM and therefore this is
an example of weakly interacting new physics. There is a
continuum of such models characterized by an angle
$0^0\le\theta_6\le 90^0$. To any value of $\theta_6$ it
corresponds a different model. The 95\% CL regions allowed by
$Q_W$, in the plane $(\theta_6,M_{Z'})$, for different values of
the Higgs mass, are shown in Fig. \ref{zetap}. In deriving these
Figures the assumption of zero mixing between $Z$ and $Z'$ has
been made. In the Figure are also shown three popular models:
$\eta$ ($\theta_6\approx -52^0$), $\chi$ ($\theta_6=0^0$), $\psi$
($\theta_6=90^0$). We see that the $\eta$ and the $\psi$ models
are not allowed by the data. The direct search at the Tevatron
for a $Z'$ within the $\chi$ model gives a direct lower bound at
95\% CL, $M_{Z'}\ge 590~GeV$ (a similar bound holds for all these
models) . Therefore this model is compatible with  the data. A
recent best fit to all the data (including APV) gives for the
$\chi$ model the following results \cite{erler},
$M_{Z'}=812^{+339}_{-152}~GeV$ and a mixing angle compatible with
zero, $\theta_M=(-1.12\pm 0.80)\times 10^{-3}$.

\section*{Atomic physics and CPT violation}

The CPT theorem is one of the fundamental results in local
relativistic field theories. Therefore the idea of possible
violations of this theorem implies that some of the axioms of
these theories should be reviewed. Let me recall here the exact
statement of the theorem \cite{cpt}: {\it In a field theory
satisfying
\begin{enumerate}
  \item Locality
  \item Lorentz invariance
  \item Analiticity of the Lorentz group representations in the
  boost parameters
\end{enumerate}
the CPT transformation is a symmetry of the theory itself}.

The first two conditions say that one is dealing with a local
relativistic field theory, whereas the third one is satisfied in
any finite-dimensional representation of the Lorentz group. It is
interesting to notice that unitary representations fail to be
analytic and as a consequence the CPT theorem can be violated in
this case. The first example of this situation dates back to
Majorana \cite{majorana} when he formulated a first order wave
equation without negative-energy solutions. He was able to do that
by making use of a unitary infinite-dimensional representation of
the Lorentz group. Since this theory does not contain
antiparticles the CPT symmetry is broken. However, the quarks and
leptons described by the SM belong to finite-dimensional
representation of the Lorentz group and therefore this does not
seem a possible way to break the theorem. It seems also very hard
to give up  locality, since it guarantees the microcausality of
the theory. Therefore, the only  sensible way to avoid the
consequences of the CPT theorem in a local field theory  seems to
break  Lorentz invariance. A situation of this type could arise at
a more fundamental level as in string theory, where it is possible
that Lorentz invariance is spontaneously broken around the Planck
mass, $M_P$ \cite{kostelecky1}. One can take into account these
effects by writing down a local effective lagrangian with Lorentz
and CPT breaking terms. These terms can be written as an
expansion in derivatives over the Planck mass. For instance,
considering a single fermion, the violating term can be written as
\be {\cal L}_v=\sum_{n}\frac{g_n}{M_P^n}
T\bar\psi\Gamma(i\partial)^n\psi\ee I have used a somewhat
symbolic notation where $\Gamma$ stays for a generic combination
of Dirac matrices and $T$ is a constant tensor and I take the
mass dimensions of $g_n$ as $[g_n]=1$. Furthermore I will assume
the same internal symmetries as in the SM, that is $SU(3)\otimes
SU(2)\otimes U(1)$ \cite{kostelecky2,kostelecky3}. Since the
breaking terms should vanish in the limit $M_P\to\infty$ also for
$n=0$, I will require \be g_0=c_o\frac{m^2}{M_P}\ee where $m$ is
some low-energy mass scale parameter. We see that the relevant
terms are the ones with $n=0$ and $n=1$, and therefore the
resulting theory preserves the renormalizability property.

Let me now consider a single fermion interacting with the
electromagnetic field. One adds to the standard QED lagrangian the
following two terms \be {\cal
L}_v^{(n=0)}=\bar\psi[-a_\mu\gamma^\mu-b_\mu\gamma_5\gamma^\mu-\frac
12 H_{\mu\nu}\sigma^{\mu\nu}]\psi\label{L0}\ee and \be {\cal
L}_v^{(n=1)}=\bar\psi[ic_{\mu\nu}\gamma^\mu
D^\nu+id_{\mu\nu}\gamma_5\gamma^\mu D^\nu]\psi\label{L1}\ee where
$D_\mu=\partial_\mu-iq A_\mu$, with $q$ the electric charge of
the fermion. There are other possible terms with $n=1$, but they
are not compatible with the symmetries of the SM and therefore
they should be suppressed. The following orders of magnitude are
expected  \be a_\mu, b_\mu, H_{\mu\nu}\approx {\cal
O}\,(m^2/M_P),~~~c_{\mu\nu}, d_{\mu\nu}\approx {\cal
O}\,(m/M_P)\label{tensors}\ee The terms in ${\cal L}_v^{(n=0,1)}$
violate Lorentz invariance, since all the tensors in eq.
(\ref{tensors}) are constant ones. However only the terms
proportional to $a_\mu$ and $b_\mu$ violate CPT symmetry since
$\gamma_\mu$, $\gamma_\mu\gamma_5$ and $D_\mu$ are CPT odd,
whereas  the other covariant terms are CPT even. Therefore, in
the following I will take into consideration only ${\cal
L}_v^{(n=0)}$. Notice also that when dealing with a single
fermion the term in $a_\mu$ does not have physical meaning since
we can write $a_\mu=\partial_\mu(a\cdot x)$, showing that $a_\mu$
is a trivial gauge background field. Of course, the situation
changes when dealing with different fermions having different
$a_\mu$'s. From eq. (\ref{tensors}) we expect that the order of
magnitude of the CPT and Lorentz breaking terms is given by
$m/M_P\approx 10^{-22}\div 10^{-17}$  for $m=m_e\div v$, where
$m_e$ is the electron mass and $v\approx 250~GeV$ is the
electroweak symmetry breaking scale. Lorentz and CPT breaking
terms could appear also in the photon part of the total
lagrangian. This instance is discussed thoroughly in the second
paper of ref. \cite{kostelecky2}, but I will not consider it in
this talk.

Here I want to illustrate   some  atomic physics experiment about
CPT violation. But before doing that let me just give a list of
other existing or planned experiments about the violation of this
fundamental symmetry
\begin{itemize}
  \item $K-\bar K$ mass difference. This experiment gives the best
  high-energy result \cite{PDG}
  \be \frac{|m_K-m_{\bar K}|}{m_K}\lesssim 10^{-18}\label{meson}\ee
  \item Experiments on neutral meson oscillations to be done at
  meson factories \cite{kostelecky4}.
  \item Experiments on muons \cite{kostelecky5}.
  \item Experiments with spin-polarized solids \cite{bluhm1}.
  \item Experiments from clock-comparison \cite{kostelecky6}.
\end{itemize}
CPT violation may have also some relevance for baryogenesis and
this subject has been  discussed in ref. \cite{bertolami}.

Let me now consider atomic physics experiments for testing CPT
using atomic traps. Several of these experiments have been
performed by confining single particles or antiparticles in a
Penning trap for a long time. These experiments have a very high
precision, of order $10^{-9}$ or better, whereas the precision in
experiments about mesons (see eq. (\ref{meson})) is much lower,
of order $10^{-3}$. I recall here the comparison of the electron
and positron gyromagnetic ratios, $g_{\mp}$, obtained measuring
their cyclotron and anomaly frequencies (see later),  which gives
the figure of merit \cite{dyck}\be \left |\frac{g_--g_+}{g_{\rm
av}}\right |\lesssim 2\times 10^{-12}\ee  Measuring the proton
and antiproton cyclotron frequencies, one can get their
charge-to-mass ratios. $r_{p.\bar p}$ \cite{gabrielse} \be\left
|\frac{r_p-r_{\bar p}}{r_{\rm av}}\right |\lesssim 9\times
10^{-11}\ee Analogously, from the charge-to-mass ratio for
electron and positron \cite{schwinberg}\be\left
|\frac{r_{e^-}-r_{e^+}}{r_{\rm av}}\right |\lesssim 1.3\times
10^{-7}\ee As we see the relevant figures of merit are much
bigger than the one for the mass difference $K-\bar K$, although,
as noticed, these measurements are about six order of magnitude
more sensitivity than the one leading to (\ref{meson}). In ref.
\cite{bluhm2} it has been argued that these figures of merit
could not be the relevant ones in testing CPT breaking. In fact,
within the approach presented here, at the lowest order in the
CPT violating parameters, one has $g_-=g_+$, and similarly the
charge-to-mass ratios do not depend on these parameters
\cite{bluhm2}. To review this point, let me start by the Dirac
equation for an electron or a proton including the breaking terms
contained in ${\cal L}_v^{(n=0)}$ (of course, the breaking
parameters may depend on the type of particle one is considering)
\be\left(i\gamma^\mu D_\mu-m-b_\mu\gamma_5\gamma^\mu-\frac 1 2
H_{\mu\nu}\sigma^{\mu\nu}\right)\psi=0 \ee In a Penning trap the
radial confinement is obtained through a strong axial magnetic
field, whereas the axial confinement is obtained by a quadrupole
electric field. The main corrections due to the CPT and Lorentz
breaking parameters are obtained by taking $A_\mu$ as the
four-potential for a constant magnetic field. Then, to obtain the
energy shifts generated by the breaking parameters one makes use
of the relativistic Landau levels wave functions and the
expressions containing the full QED corrections for the
unperturbed levels \cite{bluhm2,bluhm3}. However, the underlying
physics can be understood quite simply recalling the expression
for the non-relativistic Landau levels \be
E_{n,\sigma}=\left(n+\frac 1 2+\frac g 2\right)\frac{Be}
m,~~~\sigma=\pm \frac 1 2\ee The cyclotron and anomalous
frequencies are obtained comparing two Landau levels with
different quantum number $n$ and with the same and opposite spin
configurations respectively \bea
\omega_c&=&E_{1,-1/2}-E_{0,-1/2}=\frac{Be}m\nn\\
\omega_a&=&E_{0,+1/2}-E_{1,-1/2}=\frac{g-2}2\frac{Be}m\eea The
relevant CPT and Lorentz breaking corrections to the energy
levels are given by \cite{bluhm3} \be\delta E^{e^-}_{n,\pm 1/2}
=\mp b_3\pm H_{12},~~~\delta E^{e^+}_{n,\pm 1/2} =\mp b_3\mp
H_{12}\ee where we have taken the third axis along the magnetic
field of the trap. The frequencies for the antiparticles that we
need according to the CPT theorem are the ones with inverted spin,
therefore \be \omega_c^{e^-}=\omega_c^{e^+}=\omega_c,~~~ To
conclude thisomega_a^{e^\mp}=\omega_a\mp 2 b_3+2H_{12}\ee We get
\be\Delta\omega_c\equiv
\omega_c^{e^-}-\omega_c^{e^+}=0,~~~\Delta\omega_a\equiv\omega_a^{e^-}
-\omega_a^{e^+}=-4b_3\ee We recall that these equations hold only
at the first order in the breaking parameters and also that the
usual relation $(g-2)/2=\omega_a/\omega_c$ does not hold here
since, as noted before, the gyromagnetic ratios do not change at
the lowest order.

Since the observables that are measured in a Penning trap are the
anomalous and cyclotron frequencies, it seems natural to
introduce figures of merit related to these observables. A such
figure of merit for CPT violation is \cite{bluhm2} \be
r^e_{\omega_a}=\frac{|{\cal E}^{e^-}_{n,\sigma}-{\cal
E}^{e^+}_{n,-\sigma}|}{{\cal E}^{e^-}_{n,\sigma}}=\frac{|\delta
E^{e^-}_{n,\sigma}-\delta E^{e^+}_{n,-\sigma}|}{{\cal
E}^{e^-}_{n,\sigma}} \ee where ${\cal E}=E+\delta E$. For a weak
magnetic field one gets \be
r^e_{\omega_a}=\frac{|\Delta\omega_a|}{2m}=2\frac{|b_3|}m\ee A
new analysis of the 1987 experiment by Dehmelt et al. \cite{dyck}
has been done recently in ref. \cite{dehmelt}  obtaining the
following bound  \be r^e_{\omega_a}\lesssim 1.2\times 10^{-21}\ee
However, the vector $b_\mu$ is absolutely constant and as such it
rotates with a diurnal period of 23 h and 56 m, when seen in the
laboratory frame wich is fixed with respect to the earth. This
effect might have given rise to non favorable situations during
the observation, and therefore the bound has been a bit relaxed
\cite{dehmelt} \be r^e_{\omega_a}\lesssim 3\times 10^{-21}\div
2\times 10^{-20}\ee In the case of proton and atiproton there is
no experiment at the moment. Assuming an experimental sensitivity
analogous to the electron positron case (meaning
$\delta\omega_a\approx 2~Hz$) one gets \cite{bluhm3} \be
r^p_{\omega_a}=2\frac{|b_3^p|}m_p\lesssim   10^{-23}\ee

The last case I consider is the spectroscopy of free or
magnetically trapped  hydrogen ($H$) and antihydrogen ($\bar H$).
This is interesting since  the two-photon $1S-2S$ transition has
been measured with a precision of $3.4\times 10^{-14}$
\cite{udem} in a cold atomic beam of $H$ and with a precision of
$10^{-12}$ in trapped $H$ \cite{cesar}. However for the free case
the dependence of the $1S-2S$ transition on the CPT and Lorentz
breaking parameters is suppressed by a factor $\alpha^2/8\pi$,
since  the $1S$ and $2S$ levels shift by the same amount at the
leading order  in the breaking \cite{bluhm4}. Consider now the
spectroscopy of $H$ and $\bar H$ in a magnetic field $B$. In the
basis $|m_J,m_I\rangle$ the four $1S$ and $2S$ hyperfine Zeeman
levels are, for $n=1,2$ \bea
&&|b_n\rangle=|-1/2,-1/2\rangle,~~~|d_n\rangle=|1/2,1/2\rangle\nn\\
&&|a_n\rangle=\cos\theta_n|-1/2,1/2\rangle-\sin\theta_n
|1/2,-1/2\rangle\nn\\
&&|c_n\rangle=\sin\theta_n|-1/2,1/2\rangle+\cos\theta_n|1/2,-1/2\rangle\eea
with $\tan 2\theta_n=(51~{\rm mT})/n^3 B$.  Transitions of the
type $|c_1\rangle\to|c_2\rangle$ have leading-order sensitivity
to Lorentz and CPT violation, but they are field-dependent. As a
consequence there is a problem connected with the broadening of
the lines due  to trapping field inhomogeneities.

Consider now hyperfine transitions in the ground state. Again
there is the problem of the Zeeman broadening.  However one can
try to eliminate the frequency dependence on $B$ (at lowest
order) by choosing a field independent transition point
\cite{bluhm4}. For $B\approx 0.65~T$ the state $|c_1\rangle$ is
highly polarized ($|1/2,-1/2\rangle$). Then the effect on the
transition $|c_1\rangle\to |d_1\rangle$ of the CPT and Lorentz
violating parameters is $\delta\omega_{c\to d}^{H,\bar H}=2(\mp
b_3^p+H_{12}^p)$. Therefore by putting $ \Delta\omega_{c\to
d}=\omega^H_{c\to d}-\omega^{\bar H}_{c\to d}$ the corresponding
figure of merit can be defined as \be r_{c\to
d}^H=\frac{|\Delta\omega_{c\to d}|}{m_H}=4\frac{|b_3^p|}{m_H}\ee
Attaining a resolution of $1~mHz$, one would get \cite{bluhm4} \be
r_{c\to d}^H\lesssim 5\times 10^{-27}\ee

\section*{Conclusions}

In this talk I have reviewed some important consequences of
atomic physics measurements in the domain of high-energy physics.
In particular APV in cesium could be the  first real indication of
new physics beyond the SM. The atomic physics tests of the CPT
symmetry are already  at a spectacular level of sensitivity, and
the future experiments on $H$ and $\bar H$ could give bounds well
below the one expected from string theory.


\newpage

\begin{references}
\bibitem{bouchiat2} Bouchiat M.A. and Bouchiat C.C., {\it Phys. Lett.} {\bf
B48}, 111 (1974); {\it J. Phys.} {\bf 35}, 899 (1974).

\bibitem{musolf}
Ramsey-Musolf M.J., {\tt physics/0001250}, (2000).

\bibitem{bouchiat} Bouchiat M.A., Guena J., Hunter L. and Pottier L.,
{\it Phys. Lett.}\ {\bf B117}, 358 (1982).

\bibitem{boulder1} Noecker M.C., Masterson B.P. and  Wieman C.E.,
{\it Phys. Rev. Lett.} {\bf 61}, 310 (1988).

\bibitem{boulder2}
Wood C.S., Bennett S.C., Cho D., Masterson B.P., Roberts J.L.,
Tanner C.E. and Wieman C.E., {\it Science} {\bf 275}, 1759 (1999).

\bibitem{thallium} Vetter P.A., Meekhof D.M., Majumder P.K.,
Lamoreaux S.K. and Fortson E.N., {\it Phys. Rev. Lett.} {\bf 71},
3442 (1993); Edwards N.H., Phipp S.J., Baird E.G., Nakayama S.,
{\it Phys. Rev. Lett.} {\bf 74}, 2654 (1995).

\bibitem{dzuba} Dzuba V.A., Flambaum V.V., Silvestrov P. and Sushkov O.,
{\it Phys. Lett.} {\bf A141}, 147 (1989).

\bibitem{blundell} Blundell S.A., Johnson W.R. and Sapirstein J.,
{\it Phys. Rev. Lett.} {\bf 65}, 1411 (1990).

\bibitem{bennett}
Bennett S.C. and Wieman C.E., {\it Phys. Rev. Lett.} {\bf 82},
2484 (1999).

\bibitem{dzuba2} Dzuba V.A., Flambaum V.V. and Ginges J.S.M.,
contribution A11 to this meeting, {\it Conference Abstracts}, eds.
F. Fusi and F. Cervelli. See also Dzuba V.A. and Flambaum V.V.,
{\tt physics/0005038}, (2000).

\bibitem{pollock}
Pollock S.J. and Welliver M.C., {\it Phys. Lett.}, {\bf B464},
177 (1999).


\bibitem{derevianko}
Derevianko A. {\tt physics/0001046}, (2000); Derevianko A., {\tt
hep-ph/0005274}, (2000).

\bibitem{kozlov}
Kozlov M.G., Porsev S.G. and Tupitsyn I.I., {\tt physics/0004076},
(2000) and contribution A15 to this meeting, {\it Conference
Abstracts}, eds. F. Fusi and F. Cervelli.



\bibitem{marciano}
 Marciano W.J. and  Rosner J.L., {\it Phys. Rev. Lett.} {\bf 65}, 2963 (1990);
  Altarelli G., Lectures given at the {\it Les Houches Summer
School: Particles In The Nineties}, 30 Jun - 26 Jul 1991, Les
Houches, France.


\bibitem{altarelli}
 Altarelli G.,  Barbieri R. and  Jadach S., {\it Nucl. Phys.} {\bf B369}, 3
(1992);  Altarelli G.,  Barbieri R. and  Caravaglios F. {\it Nucl.
Phys.} {\bf B405}, 3 (1993); {\it ibidem Phys. Lett.} {\bf B349},
145 (1995).

\bibitem{peskin}
 Peskin M.E. and  Takeuchi T., {\it Phys. Rev. Lett.} {\bf 65}, 964
 (1990); {\it ibidem} {\it Phys. Rev.} {\bf D46}, 381 (1991).

\bibitem{altarelli2}
 Altarelli G.,  Barbieri R. and  Caravaglios F., {\it Int. J. Mod.
 Phys.} {\bf A13}, 1031 (1998), and updating of these results as
 communicated to me by Dr. Caravaglios.

\bibitem{higgs}
The actual experimental lower bound on the Higgs mass is 107.9
$GeV$ at 95\% CL, see: ALEPH, DELPHI, L3 and Opal Collaborations.
The LEP Working group for Higgs boson searches, CERN-EP-2000-055,
April 2000.



\bibitem{chen}
Pollock S.J., Fortson E.N. and Wilets L., {\it Phys. Rev.} {\bf
C46}, 2587 (1992); Chen B.Q. and Vogel P., {\it Phys. Rev.} {\bf
C48}, 1392 (1993).

\bibitem{casalbuoni}
Casalbuoni R., De Curtis S., Dominici D. and  Gatto R., {\it Phys.
Lett.} {\bf B460}, 135 (1999).

\bibitem{langacker}
Langacker P., {\it Phys. Lett.} {\bf B256}, 277, (1991).

\bibitem{PDG}
Particle Data Group, Caso C. {\it et al.}, {\it Eur. Phys. J.}
{\bf C3}, 1 (1998).

\bibitem{pomarol}
Pomarol A. and Quiros M., {\it Phys. Lett.} {\bf B438}, 255
(1998); Delgado A., Pomarol A. and Quiros M., {\it Phys. Rev.}
{\bf D60}, 95008 (1999); Masip M. and Pomarol A. {\it Phys. Rev.}
{\bf D60}, 96005 (1999).


\bibitem{casalbuoni2}
 Casalbuoni R.,  De Curtis S.,  Dominici D. and  Gatto R., {\it Phys. Lett.}
 {\bf B462}, 48 (1999).

\bibitem{qwz}
Amaldi U. {\it et al.}, {\it Phys. Rev.} {\bf D36}, 1385 (1987);
Marciano W.J. and Rosner J.L., {\it Phys. Rev. Lett.} {\bf 65},
2963 (1990); Altarelli G. {\it et al.} {\it Phys. Lett.} {\bf
B261}, 146 (1991);  Mahantappa K.T. and Mohapatra P.K., {\it
Phys. Rev} {\bf D43} 3093 (1991); Rosner J.L., {\it Phys. Rev.}
{\bf D61}, 016006 (2000). In these papers it is also possible to
find a complete list of references to the $Z'$ models.

\bibitem{erler}
Erler J. and Langacker P., {\it Phys. Rev. Lett.} {\bf 84}, 212
(2000).

\bibitem{cpt}
Streater R.F. and Wightman A.S., {\it PCT, spin and statistics
and all that}, edited by W.A. Benjamin, Inc., New York,
Amsterdam, 1964.

\bibitem{majorana}
Majorana, E., {\it Il Nuovo Cimento} {\bf 9}, 335 (1932). This
paper is in italian and it was translated in english by Fradkin
E.S., {\it AJP} {\bf 34}, 314 (1966).

\bibitem{kostelecky1}
Kosteleck\'y V.A. and  Potting R., {\it Nucl. Phys.} {\bf B359},
545, (1991); {\it ibidem}, {\it Phys. Lett.} {\bf B381}, 389,
(1996);  Kosteleck\'y V.A. and  Samuel S., {\it Phys. Rev. Lett.}
{\bf 63}, 224, (1989); {\it ibidem}, {\it Phys. Rev. Lett.} {\bf
66}, 1811, (1991); {\it ibidem}, {\it Phys. Rev.} {\bf D39}, 683,
(1989); {\it ibidem}, {\it Phys. Rev.} {\bf D40}, 1886, (1989).

\bibitem{kostelecky2}
Colladay D. and Kosteleck\'y V.A., {\it Phys. Rev.} {\bf D55},
6760, (1997); {\it ibidem}, {\bf D58}, 116002, (1998).

\bibitem{kostelecky3}
For a recent review of the results on this subject, see
Kosteleck\'y V.A., {\tt hep-ph/0005280}, (2000).

\bibitem{kostelecky4}
Kosteleck\'y V.A., {\it Phys. Rev. Lett.} {\bf 80}, 1818, (1998).

\bibitem{kostelecky5}
Bluhm R., Kosteleck\'y V.A. and Lane C.D., {\it Phys. Rev. Lett.}
{\bf 84}, 1098, (2000).

\bibitem{bluhm1}
Bluhm R. and Kosteleck\'y V.A., {\it Phys. Rev. Lett.} {\bf 84},
1381, (2000).

\bibitem{kostelecky6}
Kosteleck\'y V.A. and Lane C.D., {\it Phys. Rev.} {\bf D60},
116010, (1999).

\bibitem{bertolami}
Bertolami O. {\it et al.}, {\it Phys. Lett.} {\bf B395}, 178
(1997).

\bibitem{dyck}
Van Dyck R.S. Jr., Schwinberg P.B. and Dehmelt H.G., {\it Phys.
Rev. Lett.} {\bf 59}, 26 (1987); {\it ibidem} {\it Phys. Rev.}
{\bf D34}, 722 (1986). For a review of the principles of the
Penning trap, see: Brown L.S. and Gabrielse G. {\it Rev. Mod.
Phys.} {\bf 58}, 233 (1986).

\bibitem{gabrielse}
Gabrielse G. {\it et al.}, {\it Phys. Rev. Lett.} {\bf 82}, 3198
(1999).

\bibitem{schwinberg}
 Schwinberg P.B., Dyck R.S. Jr.  and Dehmelt H.G., {\it Phys.
Lett.} {\bf A81}, 119 (1981).


\bibitem{bluhm2}
Bluhm R.,  Kosteleck\'y V.A. and Russell N., {\it Phys. Rev.
Lett.} {\bf 79}, 1432 (1997).

\bibitem{bluhm3}
Bluhm R., Kosteleck\'y V.A. and Russell N., {\it Phys. Rev.} {\bf
D57}, 3932 (1998).

\bibitem{dehmelt}
Dehmelt H., Mittleman R., Van Dyck R.S. Jr. and  Schwinberg P.,
{\it Phys. Rev. Lett.} {\bf 83}, 4694 (1999).

\bibitem{udem}
Udem T. {\it et al.}, {\it Phys. Rev. Lett.} {\bf 79}, 2646
(1997).

\bibitem{cesar}
Cesar C.L. {\it et al.}, {\it Phys. Rev. Lett.} {\bf 77}, 255
(1996).

\bibitem{bluhm4}
Bluhm R.,  Kosteleck\'y V.A. and Russell N., {\it Phys. Rev.
Lett.} {\bf 82}, 2254 (1999).

\end{references}
\end{document}